\documentclass[conference]{IEEEtran}

\usepackage{graphicx}
\usepackage{amssymb}
\usepackage{subfig}
\usepackage{pdflscape}
\usepackage{caption}

\begin{document}
\title{Compressed EEG Acquisition with Limited Channels using Estimated Signal Correlation }

\author{\IEEEauthorblockN{J V Satyanarayana, Member IEEE}
\IEEEauthorblockA{Department of Electrical Engineering\\
Indian Institute of Science\\
Bangalore, 560012\\
Email: jvsat29@yahoo.co.in}
\and
\IEEEauthorblockN{A G Ramakrishnan, Senior Member IEEE}
\IEEEauthorblockA{Department of Electrical Engineering\\
Indian Institute of Science\\
Bangalore, 560012\\
Email: ramkiag@ee.iisc.ernet.in}
%\IEEEauthorblockN{Homer Simpson}
%\IEEEauthorblockA{Twentieth Century Fox\\
}

\maketitle

\begin{abstract}
\textbf{Nearby scalp channels in multi-channel EEG data exhibit high correlation. A question that naturally arises is whether it is required to record signals from all the electrodes in a group of closely spaced  electrodes in a typical measurement setup. One could save on the number of channels that are recorded, if it were possible to reconstruct the omitted channels to the accuracy needed for identifying the relevant information (say, spectral content in the signal), required to carry out a preliminary diagnosis. We address this problem from a compressed sensing perspective and propose a measurement and reconstruction scheme.  Working with publicly available EEG database, we put our scheme to experiment and illustrate that if it is only a matter of estimating the frequency content of the signal in various EEG bands, then all the channels need not be recorded. We have achieved an average error below 15\% between the original and reconstructed signals with respect to estimation of the spectral content in the delta, theta and alpha bands. We have demonstrated that channels in the 10-10 system of electrode placement can be estimated, with an error less than 10\% using recordings on the sparser 10-20 system.}\\
\end{abstract}

\begin{keywords}
\textbf{Correlated signals, Karhunen-Loeve Transform, Electroencephalography, Motor-Imagery tasks, Compressed Sensing,  Convex Optimization, EEG electrode placement}
\end{keywords}

% \fi
\IEEEpeerreviewmaketitle

\section{Introduction}
\label{sec:intro}
%{Electroencephalography \cite{eegsp,eegbasic} is the neurophysiological measurement of the electrical activity of the brain using electrodes placed on the scalp. The resulting traces are known as electroencephalogram (EEG) and they represent an electrical signal (postsynaptic potentials) from a large number of neurons. The EEG is a brain non-invasive procedure frequently used for diagnostical purposes. Instead of electrical currents the voltage differences between different parts of the brain are observed.  Internationally standardized systems like the 10-20 system and the 10-10 system are usually employed to record the spontaneous EEG. For example, in the 10-20 system, 21 electrodes are located on the surface of the scalp. EEG measurements allow both time-domain and frequency-frequency analysis. The following important frequency bands have been identified in the EEG signal:
%\begin{itemize}
  %\item delta - 3 Hz and less (deep sleep, when awake pathological)
	%\item theta - 3.5 - 7.5 Hz (creativity, falling asleep)
	%\item alpha - 8 - 13 Hz (relaxation, closed eyes)
	%\item beta - 14 - 30 Hz and more (concentration, logical and analytical thinking, fidget)
	%\item gamma - greater than 30 Hz (simultaneous processes)
%\end{itemize}

Nearby scalp channels of multi-channel EEG exhibit high correlation because EEG signals are not produced in the scalp or the neurons (brain) directly under the recording electrodes. Instead, as suggested in \cite{eegica}, they are a consequence of partial synchrony of local field potentials from distinct cortical domains - each domain, in the simplest case,  being a patch of cortex of unknown extent. At any electrode, the EEG recording is a weighted linear mixture of underlying cortical source signals. The strong correlations observed between EEG recordings at nearby electrodes can be attributed to the spatial mixing of EEG source signals by volume conduction. Significant research effort has gone into exploring  the correlation between EEG recordings at electrodes on different areas of the scalp. In \cite{eegcoh,sourceloc,cortimage}, heavy correlation, sometimes as high as 0.9 has been reported between anterior-posterior EEG signals in the alpha band. Very high coherence in the delta band  has been reported in \cite{scizh} between posterior temporal lobe regions. Interhemispheric coherence in the gamma band has been studied in normal adults in \cite{interhem}. Existence of very high correlation, between temporal regions of the human brain, in the alpha band has been reported in \cite{alphacorr}. 

\subsection{Motivation}
\label{subsec:mot}
The primary interest in understanding inter-channel correlation in multi-channel EEG is to identify scope for information redundancy in a measurement involving the full set of electrodes. If the channels are correlated, is it always mandatory to make recordings at all the electrodes, particularly those in close vicinity ?  For any subject, during the initial training sessions, all the channels are monitored.  Once the correlation is learnt,  one could do away with measurement at some of the electrodes and yet be able to estimate the EEG spectral signature at the locations of the missing electrodes. On the other hand, it may also be possible to reconstruct channels that are noisy or missing altogether due to electrode movement, etc. This conjecture is motivated by supporting literature cited above that provides evidence of correlation between nearby EEG channels.  
\subsection{Limitations}
\label{subsec:limit}
 One cannot ignore the fact that the inter-channel correlation may be weak and time varying.  However, in applications such as EEG brain mapping, where we are mainly concerned with the relative signal content in various frequency bands, accuracy of signal reconstruction per se, at such locations can be relaxed. The focus of this work is restricted to providing empirical evidence of obtaining approximate signal reconstruction and a reasonably good estimate of the spectral content in all channels of EEG by recording over limited number of channels. Some of the possible implications of such an effort on clinical diagnosis and research are listed in section \ref{sec:appl}. It is definitely desirable to be able to detect small changes in the correlation pattern, which might have diagnostic significance. However, we are not sure that the proposed method can accomplish this and we shall explore this in future.

\section{Sub-sampling and reconstruction}
          Reduction in the number of EEG recordings involves identification of a suitable subsampling/reconstruction architecture, realization of which based on many different paradigms has been an important research area in itself. Almost all the subsampling schemes proposed for general signals are based on the assumption of signal sparsity in some domain such as time, frequency or space. Many spectral estimation methods have been proposed, where the signal is assumed to be sparse in the frequency domain. These methods \cite{dsa,specanal,msa} are suited to applications like radar, in which targets act as spatially sparse monotones. Though EEG signals exhibit characteristic spectral peaks for various normal and pathological conditions, they are not in general sparse in the frequency domain. It is at best possible to exploit the intersignal correlation in such cases so that a vector of measurements from a set of electrodes could be transformed into a sparse vector on a basis derived from the signal autocorrelation matrix.  It is important to mention here that significant work has been reported previously on the frequency analysis of EEG signals. Time-frequency analysis of EEG data based on adaptive periodogram technique has been proposed in \cite{apt}. Identification of the signal components through decomposition of data into time-frequency-space atoms (based on the Wigner-Ville distribution) using parallel factor analysis has been proposed in \cite{parafac}. Time-frequency spectral estimation of multichannel EEG has been reported \cite{slex} using smooth, time-frequency localized versions of the Fourier functions. These contributions address the problem of detection and analysis very well. \\
          In \cite{gabor}, the authors have exploited the joint sparsity of EEG signals on the Gabor frame and have achieved a low normalized mean square error. However in this work, the different trials are treated as different electrodes with the assumption that in both cases the same underlying activity is measured. In \cite{cmos}, the authors have suggested a novel approach of structuring individual signals into groups and exploiting the group sparsity by computing the $l_{12}$ norm. However their approach involves the use of the unconventional random sampling based acquisition architecture and does not exploit the joint sparsity of a group of signals. In \cite{fica}, the authors have demonstrated the use of fast ICA as a  preprocessing step  before compressed acquisition of EEG signals to achieve a low reconstruction error. \\
          In the approach that we report in this paper, we intend to exploit the inter-channel correlation in EEG and in the process do away with some of the channels altogether during acquisition. The focus of our efforts is more on detecting the signal content in various spectral bands using reduced number of channel measurements. Section \ref{sec:klt} gives a brief introduction to the well known Karhunen Loeve Transform (KLT), which makes available a sparsifying basis for a set of correlated signals. With such a sparsifying basis, a subsampling scheme like compressed sensing could be employed for undersampling and reconstruction of the channels. The next section presents a rudimentary introduction to the area of compressed sensing which has made a significant impact in sparse signal processing in the past decade.

\section{Compressed sensing paradigm}
\label{sec:csp}
Classically, signals are Nyquist sampled and transformed into a sparse domain, following which only the significant coefficients are retained for transmission or storage. The compressed sensing paradigm \cite{cs1,cs2,cs3,cs4} provides mechanisms for sub-sampling signals, that are sparse in an arbitrary transform domain, and subsequent reconstruction of the original signal from the sub-sampled measurement. Under compressed sensing schemes, sampling and compression are combined into a single step, so that only the required smaller number of appropriate samples are obtained through non-uniform sampling at a sub-Nyquist rate. Let $\textbf{x}\in\mathbb{R}^{N}$ be a finite length, discrete signal in the time domain which is to be sub-sampled. Assume that $\textbf{x}$ has a sparse representation in a transform domain, represented by the unitary matrix  $\Psi\in\mathbb{R}^{N\times N}$. In other words,
\begin{equation}
\label{eqn:eqntran}
\textbf{x}=\Psi \textbf{X} 
\end{equation}
  where $\textbf{X}$  is an  $N\times1$ vector that has at most $K < N$ non-zero elements, i.e. it is a $K$-sparse vector. In practice, for real world signals $\textbf{x}$, $\textbf{X}$ has at most $K$ significant elements and the rest are negligibly small. For signal compression, the negligible coefficients are set to zero and only the significant coefficients are transmitted (stored). A lossy recovery of the original signal is then obtained using (\ref{eqn:eqntran}). Although the signal is efficiently compressed, all the Nyquist samples are required initially. Instead, if $M$ linear combinations of the signals are taken by a sampling matrix, $\Phi\in\mathbb{R}^{M\times N}$ we have,
  \begin{equation}
  \label{eqn:eqnlincomb}
  \textbf{f}=\Phi\textbf{x}=\Phi{\Psi}\textbf{X}
  \end{equation}
The next step after taking the compressed measurement is to recover the original signal $\textbf{x}$, given the measurement vector \textbf{f}, the inverse transformation ${\Psi^{-1}}$,  and the measurement matrix $\Phi$. It is clear that simple linear algebra does not permit us to do so, due to the fact that the set of equations (\ref{eqn:eqnlincomb}) has more number of unknowns than equations.\par
  The earliest reconstruction algorithms were geometric, involving $l_1$ minimization techniques to find the $K$-sparse vector, $\textbf{X}$ from the measurement $\textbf{f}$. 
  \begin{eqnarray} 
  \label{eqn:eqnl1min}
   minimize\, {\left|\,\textbf{\textbf{h}}\,\right|}_{1}, \ \   subject\,\, to \  \Phi\Psi\textbf{h} = \textbf{f} \nonumber  \\
   \hat{\textbf{X}}=\textbf{h} \nonumber  \\
   \hat{\textbf{x}}=\Psi\hat{\textbf{X}}
  \end{eqnarray}

Also known as the basis pursuit \cite{basispursuit}, this method has been widely used in applications of compressed sensing. It can be extended for the case of noisy signals by altering the first line in (\ref{eqn:eqnl1min}) as:
\begin{eqnarray}
	\label{eqn:eqnl1minnoise}
	minimize\, {\left|\,\textbf{\textbf{h}}\,\right|}_{1}, \ \   subject\,\, to \  {\left\|\Phi\Psi\textbf{h}-\textbf{f}\right\|}_{l_2} \leq \epsilon
\end{eqnarray}
where $\epsilon$ is a small term bounding the amount of noise in the data and whose value may be specific to the application.\par
   $l_1$ minimization technique offers a high reconstruction accuracy. However, its complexity is $\Omega\left(N^2\right)$, making it computationally intensive for large dimensional problems. Iterative greedy algorithms have been proposed, which execute faster  compromising the reconstruction accuracy. The orthogonal matching pursuit \cite{omp}, along with its many variants \cite{stomp, randproj}, is a greedy, iterative algorithm which finds the support of the sparse vector progressively. Although the basis pursuit approach is employed by the scheme  proposed in this work for acquisition and reconstruction of EEG signals, our method can work with any reconstruction algorithm.

\begin{figure*}[ht]
	\centering
	
\subfloat[10-20 system. Courtesy: Wikipedia\label{subfig:1020}]{
	   \includegraphics[scale=0.35]{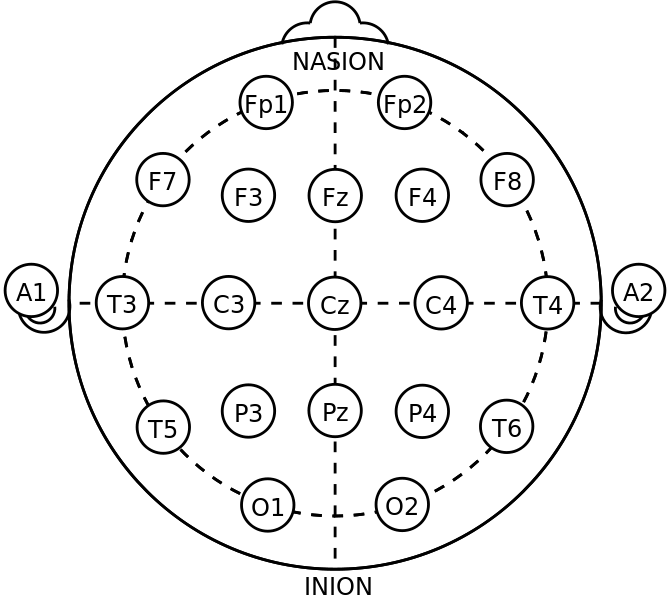}
}
%\qquad
\quad		
\subfloat[10-10 system Courtesy: Physionet\label{subfig:1010}]{
	   \includegraphics[scale=0.5]{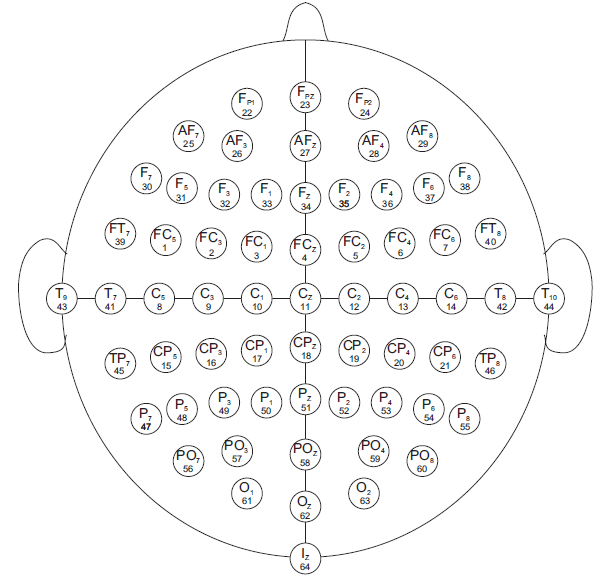}
	}
	
	\caption{EEG electrode placement systems}
	\label{fig:elect}
	
\end{figure*}

\section{Karhunen Loeve Transform}
\label{sec:klt}
Karhunen Loeve Transform (KLT) \cite{klt1, klt2} is a reversible linear transformation that removes redundancy in signals by decorrelating them. KLT has been extensively used in image compression, wherein the correlation between neighboring pixels is exploited.\par
 Consider the signal matrix, $\textbf{S}\in\mathbb{\textbf{R}}^{\tau\times N}$, the rows of which are indexed by $\tau$ successive time instants and the columns are indexed by $N$ correlated signal sources. 
The covariance matrix of \textbf{S} denoted by \textbf{$\Sigma_S$} , is symmetric for real valued \textbf{S} and its eigen vectors, $\mathbf{\psi}_n$, are orthogonal. Consequently, one can construct an orthogonal matrix, $\mathbf{\Psi\equiv\left[{\psi}_0,{\psi}_1,...,{\psi}_{N-1}\right]}$ such that \textbf{$\Sigma_S\Psi=\Psi\Lambda$} where \textbf{$\Lambda$} is a diagonal matrix consisting of the corresponding eigen values. The transpose of $\mathbf{\Psi}$ is known as the KL transform.

Let \textbf{$x^{(t)}\in{\textbf{R}}^{N\times 1}$} be a vector comprising samples from $N$ sources at any time instant $t$. Thus,  $x^{(t)}$ can be represented on the basis spanned by the eigen vectors $\mathbf{\psi}_n$ as
\begin{equation}
\label{eqn:eqnpca}
\textbf{x}^{\left(t\right)}=\mathbf{\Psi}\,\textbf{X}^{(t)}
\end{equation}

If the signals are correlated then, ${X}^{(t)}$ is going to be a sparse vector. Thus at any time instant, only a subset of the source signals need to be sampled to form a measurement vector and the remaining sources can be estimated through a suitable reconstruction scheme. Let us assume that we measure only $M$ out of the $N$ sources. Such a subsampling scheme is realized by the downsized identity matrix $\textbf{I}^{\left(M\right)}_N$ obtained by eliminating those rows from $\textbf{I}_N$ which correspond to the sources that are not measured. Thus, we have $\textbf{I}^{\left(M\right)}_N$ as the measurement matrix
\begin{eqnarray}
\phi = \textbf{I}^{\left(M\right)}_N \nonumber \\
\textbf{y}^{\left(t\right)}=\mathbf{\phi}\,\textbf{x}^{\left(t\right)}
\end{eqnarray}
where, $\textbf{y}^{\left(t\right)}$ is as before the measurement vector and from (\ref{eqn:eqnpca})
\begin{equation}
\textbf{y}^{\left(t\right)}=\mathbf{\phi}\,\mathbf{\Psi}\,\textbf{X}^{(t)}
\end{equation}
Applying the convex optimization in (\ref{eqn:eqnl1minnoise}) and the second and third equations in (\ref{eqn:eqnl1min}) we get,
\begin{equation}
   \hat{\textbf{x}}^{\left(t\right)}=\mathbf{\Psi}\hat{\textbf{X}}^{(t)}
\end{equation}
The name KLT has been synonymously used with principal component analysis (PCA) by the signal processing community. In the light of this, it is very pertinent to mention that the method proposed in this paper is not the same as the sparse PCA approach \cite{spca1,spca2} which is different from classical PCA in which the matrix $\mathbf{\Psi}$ is formed out of the eigen vectors of the autocorrelation matrix (\ref{eqn:eqnpca}). In other words, one tries to maximize $\mathbf{\psi}^{T}_{n}{\Sigma_S}{\psi}_n$ subject to $\left\|\mathbf{\psi}_n\right\|=1$.\par
 On the other hand, the sparse PCA approach seeks sparse principal components that span a low dimensional space.  The matrix $\mathbf{\Psi}$ is found by solving an optimization problem with a sparsity constraint on its entries. Equivalently, one tries to maximize $\mathbf{\psi}^{T}_{n}{\Sigma_S}{\psi}_n$ subject to $\left\|\mathbf{\psi}_n\right\|=1$ and also $\left|\mathbf{\psi}\right|=K$ where $K$ is the parameter that controls the sparsity.  As in regular PCA (or  KLT), in this paper  the sparsity constraint is not imposed on $\mathbf{\Psi}$. Instead, sparsity constraint is applied on the vector $\mathbf{X}$ in the minimization in the equations (\ref{eqn:eqnl1min}) that also involve the matrix  $\mathbf{\Psi}$ formed using the standard PCA with the help of plain matrix algebra. Thus, throughout the process of acquisition and reconstruction of the signals, the matrix $\mathbf{\Psi}$, referred to as the KLT matrix and calculated previously from the training data set, remains unaltered.
%\begin{figure*}[htbp]
	%\centering
		%\includegraphics[width=0.65\textwidth]{archi.jpg}
	%\caption{Compressed sensing architecture for acquiring super-resolved signals with relaxed specifications for AA filter}
	%\label{fig:cssuperres}
%\end{figure*}

%\begin{figure*}[htbp]
%	\centering
%		\includegraphics[width=0.57\textwidth]{filtresp.jpg}
%	\caption{Magnitude response of FIR filter of order 8}
%	\label{fig:filtresp}
%\end{figure*}

%\begin{figure*}[ht]
%	\centering
%	
%\subfloat[10-20 system. Courtesy: Wikipedia\label{subfig:1020}]{
%	   \includegraphics[scale=0.35]{10-20.png}
%}
%\qquad
%\qquad		
%\subfloat[10-10 system Courtesy: Physionet\label{subfig:1010}]{
%	   \includegraphics[scale=0.5]{64ch.jpg}
%	}
%	
%	\caption{Electrode placement systems}
%	\label{fig:elect}
%	
%\end{figure*}

\begin{table*}[htbp]
\caption{Motor/Imagery tasks during which the EEG used for the study has been collected. (see \cite{physiodesc})}
\label{tab:MotorImageryTasks}
\centering
\begin{tabular}{| l | l | l |}
\hline
           &            &  \\
Record no.       & TASK   &  DESCRIPTION OF THE TASK, DURING WHICH EEG IS RECORDED.\\ 
\hline 
           &            &  \\
Record 1   & Baseline 1 & Eyes open \textbf{DURATION: 1 sec}  \\ 
Record 2   & Baseline 2 & Eyes closed \textbf{DURATION: 1 sec}\\ 	 
Record 3   & Task 1 & A target appears on the left or the right side of the screen. The subject opens and closes \\
           &        & the corresponding fist until the target disappears. Then the subject relaxes. \textbf{DURATION: 2 sec}\\ 
          
Record 4   & Task 2 & The stimulus is same as Task 1. However, in this case the subject imagines responding \\   
           &        & to the stimulus the same way as in Task 1 and then relaxes. \textbf{DURATION: 2 sec}\\       
       
Record 5   & Task 3 & A target appears on the top or the bottom of the screen. The subject opens and closes both \\
           &        & fists if the target is on top and both feet if the target is on the bottom until the target disappears.\\
           &        & Then the subject relaxes. \textbf{DURATION: 2 sec}\\   
Record 6   & Task 4 & The stimulus is same as Task 3. Again the subject imagines responding\\
           &        & to the stimulus the same way as in Task 3 and then relaxes. \textbf{DURATION: 2 sec}\\
      \hline                                                          
\multicolumn{3}{|l|}{\textbf{The Tasks 1 to 4 are repeated two times and stored in records 7 to 14. Records 7 and 11 correspond to Task 1,}} \\
\multicolumn{3}{|l|}{\textbf{records 8 and 12 correspond to Task 2, records 9 and 13 correspond to Task 3 and records 10 and 14 correspond to Task 4.}}\\
                           
\hline     
\end{tabular}
\end{table*}

\section{Application to EEG signals}
\label{sec:eegsig}
Before exploring the possibility of applying the scheme presented in the previous section for acquisition and reconstruction of EEG signals, we present a brief introduction to the major standards of electrode placement.
\subsection{Standards for EEG electrode placement}
\label{subsec:electrode}
The first internationally accepted standard for electrode placement is the 10/20 system (figure \ref{subfig:1020}) that describes head surface locations via relative distances  over the head surface between cranial landmarks. The primary purpose of this standard is to provide a reproducible method for placing a relatively small number (typically 21) of EEG electrodes for various trials. With the advent of multi-channel EEG systems the need was felt for extending the 10/20 system to higher density electrode settings for use in research and diagnosis. This led to the introduction of the 10/10 system (figure \ref{subfig:1010}) by Chatrian et. al. in 1985, consisting of 64 electrodes, as a logical extension of the original 10/20 system. While electrodes are placed at distances of ten and twenty percent along certain contours over the scalp in the 10/20 system,  they are placed at distances of ten percent along the medial-lateral contours in the 10/10 system. Also, new contours are introduced in between the existing ones. The 10/5 system with even higher electrode density was proposed by Oostenveld and Praamstra in 2001. An elaborate description and comparison of all these systems is given in \cite{electrode}.   \par
 Consider the scenario where the electroencaphologram of a patient undergoing treatment or a subject voluntarily involved in research, has to be frequently taken. The first few sessions can constitute the training phase in which measurements from all the defined set of electrodes are taken and used to compute the inverse KLT matrix. In the subsequent sessions,  more than fifty percent of the measurements can be dispensed with thereby facilitating shorter setup time. In essence, it would be possible to employ the 10/10 system of electrodes in which signals from only fifty percent of the electrodes will be measured during sessions subsequent to the initial training phase. Thus the density of the measurements will be somewhere in between the 10/20 and 10/10 systems, whereas after the compressed reconstruction, all the signals in the 10/10 system will be available.
We have applied the proposed method to the EEG signal database from Physionet \cite{physionet}. A brief description of the database is given in the next section.

\subsection{The Physionet database}
A detailed description of the database is given in \cite{physiodesc}. This data set consists of over 1500 one- and two-minute EEG recordings, obtained from 109 volunteers. The subjects performed different motor/imagery tasks (see Table \ref{tab:MotorImageryTasks}) while 64-channel EEG was recorded, at a sampling frequency of 160 Hz,  using the BCI2000 system \cite{physioweb}.  The placement of the electrodes is as per the international 10-10 system (excluding electrodes Nz, F9, F10, FT9, FT10, A1, A2, TP9, TP10, P9, and P10). Each volunteer performed, in a sequence, a set of 14 tasks: two baseline tasks followed by four different tasks repeated three times. We have assumed that the motor/imagery tasks performed by the subjects (as given in the Table) are very benign (e.g. opening and closing of fists and imagining doing the same) and unlikely to produce ECG, EOG and EMG artifacts in the recordings. Hence, no explicit steps have been taken to filter out these signals, in case they exist.
 
\subsection{The experiments}
%In order to compare the reconstructed signals with the original, for each of the significant frequency bands - delta, theta, alpha, beta and gamma, we compute a fractional spectral measure (FSM) which is the area under the plot of the absolute value of the 512-point dft within a band as a fraction of the total plot area (0--40 Hz). For example, FSM for the \textit{delta} band is
In order to compare the reconstructed signals with the original,  we compute a fractional spectral measure (FSM) for each of the significant frequency bands - delta, theta, alpha, beta and gamma. FSM is the ratio of the sum of the absolute values of the 512-point DFT coefficients within a band to the corresponding sum in the 0--40 Hz band. Thus, FSM for the $i^{th}$ band is
\begin{equation}
FSM_{i}=\frac{\sum(abs(F_{i}))}{\sum(abs(F_{(0-40Hz)}))}
\end{equation}
where $i$  denotes one of the bands:\textit{delta, theta, alpha, beta} and \textit{gamma}.
For example, for \textit{theta} band, the numerator in the equation above is equal to the sum of the absolute values of DFT coefficients in the frequency range 4--8 Hz. We have chosen 512 as the size of the DFT so as to get a good frequency resolution of 0.3125 Hz in the DFT spectra, given that the sampling frequency of the data in the database is 160 Hz.
We compare the fractional power values for the original and the reconstructed signals. Our experiment comprises the following steps, that we categorize into the training and the testing phases for the sake of clarity:\par
   
\textbf{Training phase}\\
\textbf{Step 1}: At random,  choose six subjects to be included in the test set -- say,  1, 8, 41, 61, 77, 104\\
\textbf{Step 2}: For each subject pick a record corresponding to one of the tasks 1, 2, 3 and 4, representing different motor/imagery tasks, to be utilized for training.\\
\textbf{Step 3}: For the record that is picked, compute the inverse KLT matrix for the channels (twenty in number) - Fc5, Fc3, Fc1, Fcz, Fc2, Fc4, Fc6, C5 C3, C1, Cz, C2, C4, C6, Cp5, Cp3, Cp1, Cpz, Cp2, Cp4 on successive (non-overlapping) windows of 1000 observations each, till the end of the record. Compute the mean of the inverse KLT matrices of all the windows. The chosen channels correspond to a set of closely spaced electrodes on the scalp (see figure \ref{subfig:1010}).\\

\textbf{Testing phase}\\
\textbf{Step 4}: For each of the subjects 1, 8, 41 and 61, one of the records 7 to 10, that corresponds to the same motor/imagery task as the record which was used for computing the inverse KLT matrix, is used as the test record. For example if record 3 is used in the training phase then record 7 (that corresponds to the same motor/imagery task, see table \ref{tab:MotorImageryTasks}) is used for testing. Similarly, if the training record is 4, the testing record will be 8 and so on. In the case of subjects 77 and 104, the motor/imagery task that is chosen as the test record is different from the one used in the training phase. For example, for subject 77, records 3 and 14 are chosen for training and testing, respectively. Similarly, for subject 104, records 4 and 13 are chosen for training and testing, respectively. This is done with the objective of observing the robustness of the method.  At each successive time instant during 0-10 sec of the test record, a sub-sampling is carried out, that is, only a subset consisting of ten channels is measured. This subset is different for each subject and the members of the subset are picked up arbitrarily. The samples from the ten channels form the measurement vector, $\textbf{y}^{\left(t\right)}$. The remaining ten are estimated through $l_1$ minimization (\ref{eqn:eqnl1minnoise}) using the \textit{cvx} toolbox \cite{cvx}. The mean inverse KLT matrix is used in the compressed reconstruction algorithms.\\
\textbf{Step 5}: Compute the DFT\footnote{In order to compute the DFT, the entire signal is divided into segments of length 512, with a fifty percent overlap. The last segment is padded with sufficient number of zeros. To each segment, a 512-point Hamming window is applied. A 512-point DFT of each windowed segment is calculated. The absolute values of the DFT coefficients are averaged over all the segments.} for the original and reconstructed signals. \\
\textbf{Step 6}: Compute the fractional spectral measure for the original and the reconstructed signals and compare.\\

It is to be noted that, for any subject, all the EEG channels are sampled only during the initial training sessions. Subsequently, on the same subject, only a subset of channels need to be sampled and the rest can be reconstructed. 

\subsection{Results}
\begin{landscape}
\begin{figure}[ht]	

%	%	\subfloat{\fbox{\includegraphics[scale=0.3]{R13C01.jpg}}}\qquad	
%		\subfloat{\fbox{\includegraphics[scale=0.35]{R13C03.jpg}}}\quad
%	  \fbox{\includegraphics[scale=0.35]{R13C04.jpg}}\quad
%		\subfloat{\fbox{\includegraphics[scale=0.35]{R13C05.jpg}}} \\\\
%		\fbox{\includegraphics[scale=0.35]{R13C09.jpg}}\quad 
%		\subfloat{\fbox{\includegraphics[scale=0.35]{R13C12.jpg}}}\quad  
%		\fbox{\includegraphics[scale=0.35]{R13C13.jpg}}\\
%		\subfloat{\fbox{\includegraphics[scale=0.35]{R13C14.jpg}}}\quad 
%		\fbox{\includegraphics[scale=0.35]{R13C18.jpg}}\quad 
%		\subfloat{\fbox{\includegraphics[scale=0.35]{R13C19.jpg}}}\\\\
%	%	\fbox{\includegraphics[scale=0.3]{R13C19.jpg}}

	%	\subfloat{\fbox{\includegraphics[scale=0.3]{R13C01.jpg}}}\qquad	
		\subfloat{\fbox{\includegraphics[scale=0.45]{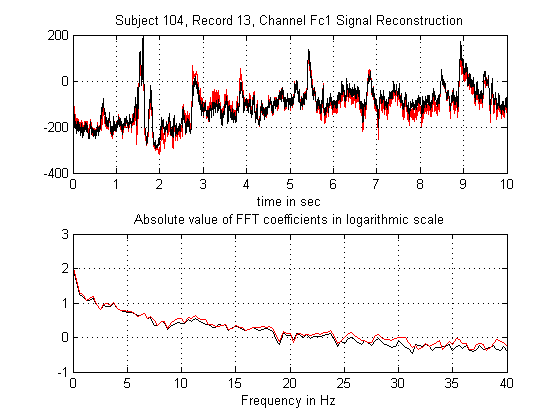}}}\qquad
	  \fbox{\includegraphics[scale=0.45]{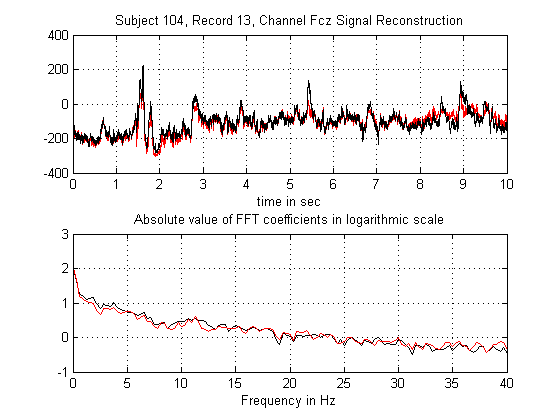}}\qquad
		\subfloat{\fbox{\includegraphics[scale=0.45]{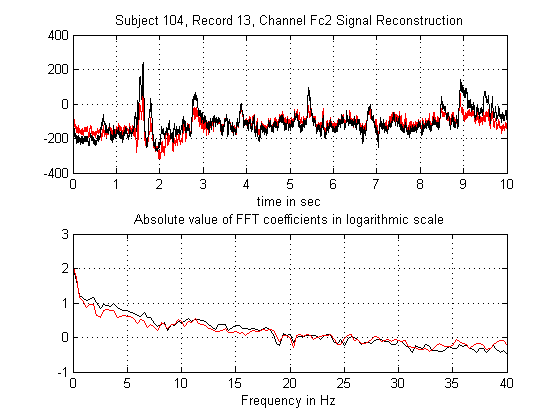}}} \\\\
		\fbox{\includegraphics[scale=0.45]{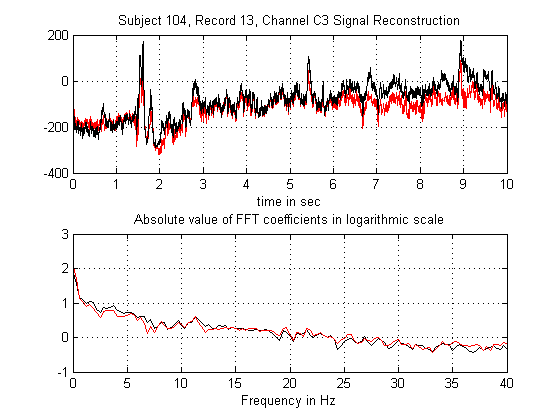}}\qquad 
		\subfloat{\fbox{\includegraphics[scale=0.45]{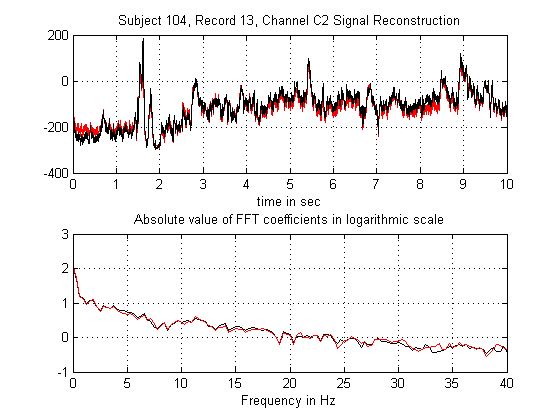}}}\qquad  
		\fbox{\includegraphics[scale=0.45]{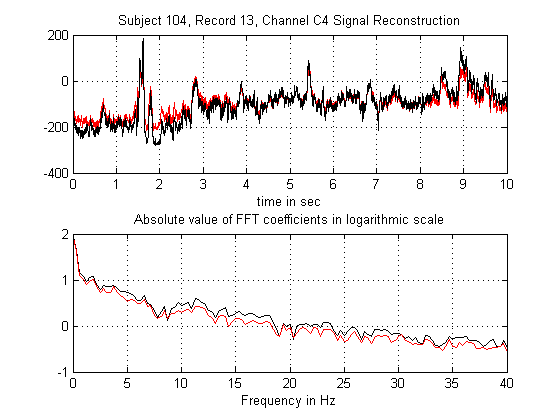}}\\
		\subfloat{\fbox{\includegraphics[scale=0.45]{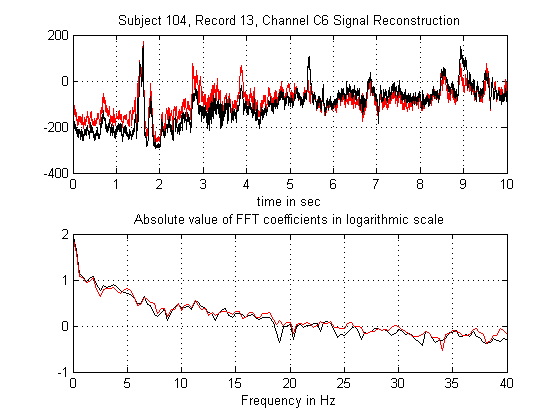}}}\qquad 
		\fbox{\includegraphics[scale=0.45]{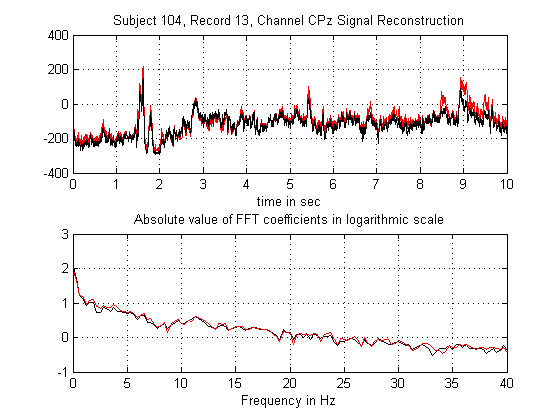}}\qquad 
		\subfloat{\fbox{\includegraphics[scale=0.45]{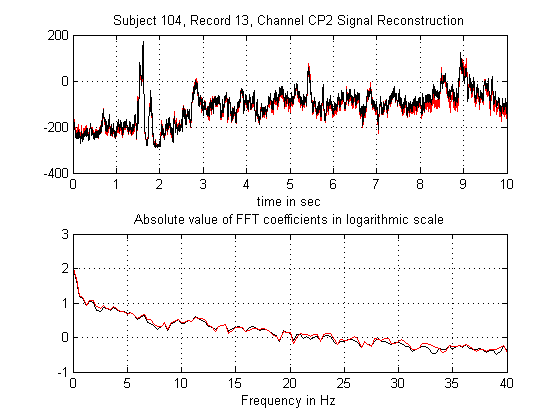}}}\\\\
	%	\fbox{\includegraphics[scale=0.3]{R13C19.jpg}}		
	\caption{Reconstructed (red) vs original (black) signals for subject 104,  record 13}
		\label{fig:s104r13}
\end{figure}
\end{landscape}

For each subject, the FSM values for each band, in each of the ten reconstructed channels are presented in Table \ref{tab:resfsm}. The average error between the original and reconstructed signals in each band is also given as a percentage.  Figure \ref{fig:s104r13} shows the plots of the reconstructed and the original signals for nine channels (only 9 out of 10 reconstructed channels are shown due to space constraint; channel 1 is not shown to facilitate symmetrical placement of the rest of the sub-figures) for subject 104.  The original and reconstructed signals have a close match in the signal as well as the frequency domains.  \par 
     
\section{Possible Applications}
\label{sec:appl}
\begin{figure*}[htbp]	
\raggedright
\centering
	%	\subfloat{\fbox{\includegraphics[scale=0.3]{R13C01.jpg}}}\qquad	
		\subfloat{\fbox{\includegraphics[scale=0.5]{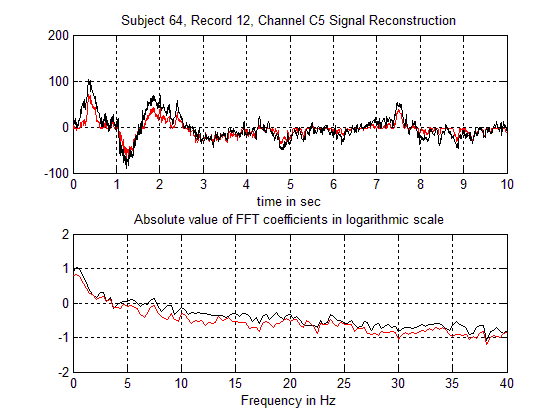}}}\quad
	  \fbox{\includegraphics[scale=0.5]{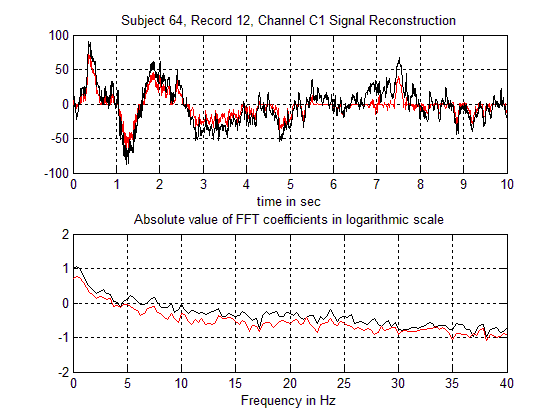}} \\   
		\subfloat{\fbox{\includegraphics[scale=0.5]{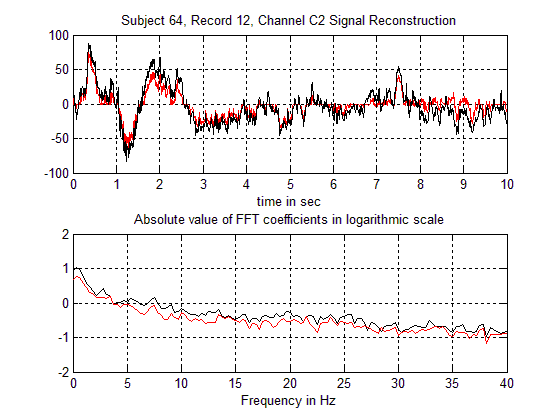}}} \quad
		\fbox{\includegraphics[scale=0.5]{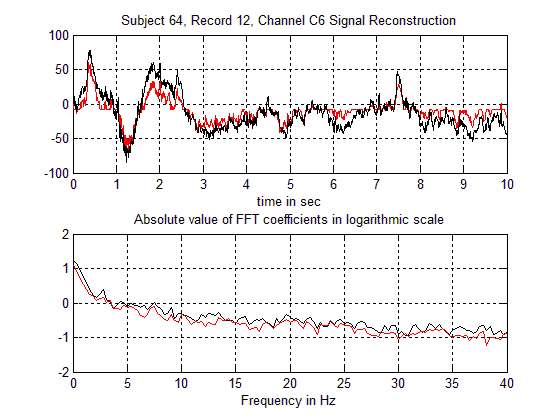}} 
				
	\caption{Estimation of 10-10 channels from 10-20 recordings - reconstructed (red) vs original (black) signals for subject 64,  record 12}
		\label{fig:1020plots}

\end{figure*}

A very pertinent question, in the context of the ideas proposed in this paper, is: are there scenarios in which after an initial training phase involving all the channels, one can dispense with some of the channels ? To answer this question, we list below a few possible cases where this could be done.
\begin{enumerate}
	\item After a one-time learning phase for any subject, it would be possible to obtain the full EEG, for the same subject, using measurements on substantially lesser number of channels. Using recordings on channels in the international 10-20 system of electrode placement, it is possible to estimate, with fair accuracy,  the spectral content of channels in the denser 10-10 system. To illustrate this, we have calculated the KLT matrix, for subject  64 in the physionet database, using channel numbers 8 to 14 of the 10-10 system, i.e. C5, C3, C1, Cz, C2, C4 and C6 that fall in a straight line on top of the scalp. For testing, using data from channel numbers 9, 11 and 13 (i.e C3, Cz and C4) that coincide with electrode locations in the 10-20 system, we reconstruct the remaining  four channels - C5, C1, C2 and C6 (see figures \ref{subfig:1020} and \ref{subfig:1010}). Table \ref{tab:rec10to20} gives the corresponding FSM values for the original and reconstructed signals. The average error in the delta band is as low as 2.6\% (or equivalently the spectral fidelity is about 97.4\%). The plots of the reconstructed channels vs the original along with the corresponding DFT magnitudes are shown in figure \ref{fig:1020plots}. \par
	 This could be useful in recording ambulatory EEG \cite{ambu} which is carried out for an extended period (up to 72 hours) in which the patient can move about freely during the recording and data is stored in a pocket recorder. Thus, EEG recorded on a subject at rest, using a dense set of electrodes, can be used for training.  Subsequently when the subject is in motion, all the channels need not be monitored. 
	\item In sleep studies, it is possible that data is missing on some channels, either due to noise or due to undesirable movement by the subject. In this case, the loss of  data, treated as undersampling, can be handled by recovery through compressed sensing. Here there is no intentional sub-sampling.
	\item Deviation between the signal values estimated through compressed sensing and the actual measurements (due to loss of correlation), beyond a threshold, can be used to detect the onset of seizure in epileptic subjects. In this case too, there is no intentional sub-sampling. 
\end{enumerate}
 \section{Conclusion and Future work}
In this work, we have presented a novel approach of subsampling EEG, by measuring only a subset of electrodes and reconstructing the remainder through compressed sensing. Empirically we have been able to demonstrate that if the correlation amongst the channels is captured with good accuracy, then by recording at only a few locations on the scalp, it is possible to estimate the relative signal content in different frequency bands with reasonable accuracy. We propose the idea that if it is a matter of only knowing the relative spectral content, measurement of only a few EEG channels suffices, provided the correlation is previously captured in the inverse KLT matrix using data reserved for training.  The method proposed is suitable for real time data capture since the computationally intensive reconstruction process can be done offline. We have demonstrated that the accuracy in estimating the relative frequency content is within 15\%. We have also shown that recordings on the 10-20 system can be used to estimate the signals on electrodes in the 10-10 system with more than 90\% spectral fidelity. Although we employed basis pursuit for the reconstruction, the approach we have presented is independent of the compressed sensing reconstruction algorithm used.\par
 For our future work, we intend to explore the possibility of detecting sudden changes such as epileptic seizures manifesting as high prediction error due to lack of correlation.

%\clearpage

\newpage
\begin{landscape}
\begin{table}[htbp]
\caption{Comparison of FSM in different bands of the original and reconstructed EEG for different channels of 6 subjects }
\label{tab:resfsm}
%\centering

\subfloat{
%\begin{tabular}{| l | l | l | l | l | l |}
\begin{tabular}{| c | c | c | c | c | c |}
\hline
\multicolumn{6}{|c|}{\textbf{Subject 1} (Record 7, Measured channels: 1 4 5 6 9 11 12 15 19 20)}\\
\hline
%& \multicolumn{2}{|c|}{Frequencies in Hz}\\
Chnl & Delta & Theta & Alpha & Beta & Gamma\\

      &    (O) \hspace{3 mm}   (R)  &    (O) \hspace{3 mm}    (R)  &    (O)\hspace{3 mm}     (R)  &   (O)\hspace{3 mm}      (R)  &   (O)\hspace{2 mm}     (R)  \\  
\hline
  Fc3  &    0.337 0.328 &  0.134 0.138 &  0.090 0.095 &  0.282 0.281 &  0.089 0.097\\
  Fc1  &    0.342 0.338 &  0.142 0.134 &  0.098 0.099 &  0.290 0.261 &  0.092 0.090\\
  Fc6  &    0.336 0.378 &  0.123 0.115 &  0.087 0.099 &  0.248 0.275 &  0.101 0.084\\
   C5  &    0.340 0.319 &  0.123 0.143 &  0.080 0.092 &  0.241 0.280 &  0.079 0.090\\
   C1  &    0.369 0.311 &  0.136 0.141 &  0.099 0.101 &  0.276 0.276 &  0.087 0.097\\
   C4  &    0.365 0.377 &  0.129 0.133 &  0.102 0.101 &  0.264 0.265 &  0.089 0.088\\
   C6  &    0.341 0.374 &  0.112 0.132 &  0.088 0.108 &  0.246 0.277 &  0.092 0.086\\
  CP3  &    0.364 0.352 &  0.142 0.128 &  0.095 0.093 &  0.272 0.256 &  0.082 0.087\\
  CP1  &    0.383 0.348 &  0.132 0.118 &  0.096 0.093 &  0.268 0.268 &  0.081 0.088\\
  CPz  &    0.384 0.351 &  0.121 0.138 &  0.096 0.105 &  0.255 0.270 &  0.079 0.092\\
\hline
Avg error & 7.2\%    &  9.0\%      &    7.7\%     &     6.2\%    &    9.4\%\\
%\multicolumn{6}{|c|}{Average error in delta band: 7.2\%}\\ %7.19
\hline
\end{tabular}
}
\qquad
\qquad
\subfloat{
%\begin{tabular}{| l | l | l | l | l | l |}
\begin{tabular}{| c | c | c | c | c | c |}
\hline
\multicolumn{6}{|c|}{\textbf{Subject 8} (Record 8, Measured channels: 1 2 6 8 9 10 13 14 16 17)}\\
\hline
%& \multicolumn{2}{|c|}{Frequencies in Hz}\\
Chnl & Delta & Theta & Alpha & Beta & Gamma\\

      &    (O) \hspace{3 mm}   (R)  &    (O) \hspace{3 mm}    (R)  &    (O)\hspace{3 mm}     (R)  &   (O)\hspace{3 mm}      (R)  &   (O)\hspace{2 mm}     (R)  \\  
\hline
  Fc1  &    0.354 0.344 &  0.146 0.124 &  0.108 0.107 &  0.200 0.211 &  0.072 0.082\\
  Fcz  &    0.370 0.354 &  0.147 0.133 &  0.104 0.108 &  0.190 0.200 &  0.073 0.074\\
  Fc2  &    0.371 0.359 &  0.147 0.141 &  0.104 0.106 &  0.189 0.195 &  0.074 0.073\\
  Fc6  &    0.346 0.323 &  0.143 0.118 &  0.101 0.099 &  0.210 0.201 &  0.093 0.084\\
   Cz  &    0.337 0.337 &  0.142 0.126 &  0.116 0.103 &  0.201 0.205 &  0.074 0.084\\
   C2  &    0.351 0.323 &  0.131 0.125 &  0.108 0.096 &  0.191 0.227 &  0.070 0.099\\
  CP5  &    0.327 0.310 &  0.116 0.123 &  0.131 0.099 &  0.208 0.254 &  0.088 0.112\\
  CPz  &    0.353 0.359 &  0.113 0.123 &  0.104 0.087 &  0.173 0.180 &  0.066 0.070\\
  CP2  &    0.344 0.320 &  0.127 0.126 &  0.116 0.116 &  0.198 0.223 &  0.073 0.100\\
  CP4  &    0.332 0.304 &  0.129 0.130 &  0.117 0.108 &  0.201 0.238 &  0.074 0.106\\
\hline
Avg error & 4.7\%    &  7.8\%      &    8.0\%     &     9.6\%    &    19.4\%\\
%\multicolumn{6}{|c|}{Average error in delta band: 4.7\%}\\ %4.712
\hline
\end{tabular}
}
\\\\
\subfloat{
%\begin{tabular}{| l | l | l | l | l | l |}
\begin{tabular}{| c | c | c | c | c | c |}
\hline
\multicolumn{6}{|c|}{\textbf{Subject 41} (Record 9,Measured channels: 4 6 7 8 11 12 15 17 19 20)}\\
\hline
%& \multicolumn{2}{|c|}{Frequencies in Hz}\\
Chnl & Delta & Theta & Alpha & Beta & Gamma\\

      &    (O) \hspace{3 mm}   (R)  &    (O) \hspace{3 mm}    (R)  &    (O)\hspace{3 mm}     (R)  &   (O)\hspace{3 mm}      (R)  &   (O)\hspace{2 mm}     (R)  \\  
\hline
  Fc5  &    0.381 0.336 &  0.163 0.160 &  0.099 0.094 &  0.221 0.248 &  0.062 0.076\\
  Fc3  &    0.447 0.324 &  0.183 0.165 &  0.079 0.094 &  0.178 0.252 &  0.049 0.078\\
  Fc1  &    0.383 0.355 &  0.185 0.172 &  0.089 0.092 &  0.238 0.240 &  0.062 0.071\\
  Fc2  &    0.370 0.345 &  0.186 0.181 &  0.090 0.091 &  0.257 0.244 &  0.063 0.071\\
   C3  &    0.336 0.351 &  0.159 0.157 &  0.098 0.083 &  0.276 0.216 &  0.073 0.070\\
   C1  &    0.348 0.348 &  0.174 0.168 &  0.101 0.092 &  0.277 0.254 &  0.070 0.074\\
   C4  &    0.339 0.342 &  0.152 0.162 &  0.085 0.085 &  0.248 0.231 &  0.065 0.064\\
   C6  &    0.372 0.360 &  0.111 0.141 &  0.066 0.072 &  0.202 0.238 &  0.062 0.071\\
  CP3  &    0.305 0.287 &  0.120 0.142 &  0.082 0.107 &  0.236 0.263 &  0.060 0.072\\
  CPz  &    0.325 0.332 &  0.152 0.149 &  0.090 0.085 &  0.255 0.230 &  0.062 0.064\\
\hline
Avg error & 7.0\%    &  8.0\%      &    9.7\%     &     13.5\%    &    15.7\%\\
%\multicolumn{6}{|c|}{Average error in delta band: 7.0\%}\\ %7.012
\hline
\end{tabular}
}
\qquad
\qquad
\subfloat{
%\begin{tabular}{| l | l | l | l | l | l |}
\begin{tabular}{| c | c | c | c | c | c |}
\hline
\multicolumn{6}{|c|}{\textbf{Subject 61} (Record 10, Measured channels: 2 6 8 9 10 14 15 17 18 19 )}\\
\hline
%& \multicolumn{2}{|c|}{Frequencies in Hz}\\
Chnl & Delta & Theta & Alpha & Beta & Gamma\\

      &    (O) \hspace{3 mm}   (R)  &    (O) \hspace{3 mm}    (R)  &    (O)\hspace{3 mm}     (R)  &   (O)\hspace{3 mm}      (R)  &   (O)\hspace{2 mm}     (R)  \\  
\hline
 	Fc5  &    0.414 0.323 &  0.137 0.132 &  0.082 0.112 &  0.219 0.267 &  0.061 0.101\\
  Fc1  &    0.408 0.349 &  0.150 0.142 &  0.086 0.089 &  0.214 0.263 &  0.068 0.090\\
  Fcz  &    0.396 0.388 &  0.140 0.125 &  0.089 0.081 &  0.209 0.225 &  0.076 0.072\\
  Fc2  &    0.390 0.392 &  0.135 0.138 &  0.081 0.086 &  0.217 0.216 &  0.101 0.075\\
  Fc6  &    0.388 0.364 &  0.135 0.133 &  0.080 0.092 &  0.216 0.242 &  0.101 0.080\\
   Cz  &    0.380 0.345 &  0.121 0.140 &  0.088 0.087 &  0.207 0.261 &  0.077 0.093\\
   C2  &    0.380 0.276 &  0.123 0.135 &  0.092 0.109 &  0.205 0.316 &  0.078 0.127\\
   C4  &    0.386 0.370 &  0.123 0.132 &  0.089 0.090 &  0.212 0.230 &  0.082 0.087\\
  CP3  &    0.375 0.236 &  0.109 0.127 &  0.084 0.114 &  0.236 0.360 &  0.067 0.127\\
  CP4  &    0.378 0.308 &  0.112 0.123 &  0.095 0.112 &  0.216 0.284 &  0.072 0.111\\
\hline
Avg error & 14.1\%    &  8.4\%      &    14.2\%     &     23.8\%    &    38.3\%\\
%\multicolumn{6}{|c|}{Average error in delta band: 14.1\%}\\%14.123
\hline
\end{tabular}
}
\\\\
\subfloat{
%\begin{tabular}{| l | l | l | l | l | l |}
\begin{tabular}{| c | c | c | c | c | c |}
\hline
\multicolumn{6}{|c|}{\textbf{Subject 77} (Record 14, Measured channels: 1 5 7 8 9 11 13 14 17 20 )}\\
\hline
%& \multicolumn{2}{|c|}{Frequencies in Hz}\\
Chnl & Delta & Theta & Alpha & Beta & Gamma\\

      &    (O) \hspace{3 mm}   (R)  &    (O) \hspace{3 mm}    (R)  &    (O)\hspace{3 mm}     (R)  &   (O)\hspace{3 mm}      (R)  &   (O)\hspace{2 mm}     (R)  \\  
\hline
  Fc3  &    0.359 0.297 &  0.171 0.140 &  0.069 0.097 &  0.144 0.182 &  0.042 0.063\\
  Fc1  &    0.359 0.353 &  0.162 0.171 &  0.063 0.096 &  0.132 0.222 &  0.040 0.073\\
  Fcz  &    0.348 0.308 &  0.162 0.149 &  0.067 0.081 &  0.139 0.191 &  0.042 0.068\\
  Fc4  &    0.376 0.283 &  0.149 0.150 &  0.060 0.100 &  0.136 0.243 &  0.038 0.071\\
   C1  &    0.347 0.285 &  0.163 0.151 &  0.069 0.104 &  0.158 0.248 &  0.045 0.082\\
   C2  &    0.336 0.346 &  0.154 0.139 &  0.068 0.089 &  0.150 0.210 &  0.040 0.070\\
  CP5  &    0.307 0.343 &  0.134 0.184 &  0.074 0.084 &  0.173 0.209 &  0.040 0.062\\
  CP3  &    0.334 0.273 &  0.148 0.152 &  0.076 0.092 &  0.181 0.240 &  0.046 0.069\\
  CPz  &    0.326 0.309 &  0.144 0.142 &  0.074 0.094 &  0.162 0.226 &  0.044 0.067\\
  CP2  &    0.336 0.279 &  0.146 0.137 &  0.073 0.086 &  0.167 0.202 &  0.041 0.060\\
\hline
Avg error & 12.9\%    &  9.8\%      &    34.0\%     &     42.0\%    &    64.0\%\\
%\multicolumn{6}{|c|}{Average error in delta band: 12.9\%}\\%12.861
\hline
\end{tabular}
}
\qquad
\qquad
\subfloat{
%\begin{tabular}{| l | l | l | l | l | l |}
\begin{tabular}{| c | c | c | c | c | c |}
\hline
\multicolumn{6}{|c|}{\textbf{Subject 104} (Record 13, Measured channels: 2 6 7 8 10 11 15 16 17 20 )}\\
\hline
%& \multicolumn{2}{|c|}{Frequencies in Hz}\\
Chnl & Delta & Theta & Alpha & Beta & Gamma\\

      &    (O) \hspace{3 mm}   (R)  &    (O) \hspace{3 mm}    (R)  &    (O)\hspace{3 mm}     (R)  &   (O)\hspace{3 mm}      (R)  &   (O)\hspace{2 mm}     (R)  \\  
\hline
   Fc5  &    0.364 0.331 &  0.138 0.095 &  0.077 0.075 &  0.175 0.171 &  0.049 0.043\\
   Fc1  &    0.363 0.360 &  0.136 0.125 &  0.079 0.085 &  0.162 0.166 &  0.036 0.041\\
   Fcz  &    0.368 0.334 &  0.134 0.123 &  0.079 0.075 &  0.158 0.168 &  0.034 0.042\\
   Fc2  &    0.360 0.327 &  0.133 0.097 &  0.078 0.082 &  0.158 0.161 &  0.033 0.045\\
    C3  &    0.345 0.317 &  0.137 0.101 &  0.087 0.078 &  0.184 0.178 &  0.042 0.043\\
    C2  &    0.349 0.348 &  0.126 0.116 &  0.082 0.079 &  0.163 0.154 &  0.034 0.037\\
    C4  &    0.346 0.358 &  0.129 0.120 &  0.086 0.076 &  0.167 0.146 &  0.036 0.034\\
    C6  &    0.350 0.318 &  0.126 0.139 &  0.078 0.084 &  0.162 0.188 &  0.043 0.052\\
   CPz  &    0.331 0.349 &  0.118 0.132 &  0.083 0.084 &  0.164 0.172 &  0.035 0.037\\
   CP2  &    0.336 0.343 &  0.123 0.121 &  0.087 0.083 &  0.169 0.162 &  0.035 0.036\\
  \hline
Avg error & 5.7\%    &  13.9\%      &    6.0\%     &     6.0\%    &    13.3\%\\
%\multicolumn{6}{|c|}{Average error in delta band: 5.7\%}\\ %5.676
\hline

%}\\
%\hline
\end{tabular}
}

\end{table}
\end{landscape}

\newpage

%\begin{landscape}
\captionsetup{width=.6\textwidth}

\begin{table*}[htbp]
\centering
\caption{Comparison (with the original) of FSM, in different bands, of the 10-10 system EEG channels reconstructed through compressed sensing using only recordings done on the 10-20 system\\  }
\label{tab:rec10to20}

%\begin{tabular}{| l | l | l | l | l | l |}
\begin{tabular}{| c | c | c | c | c | c |}
\hline
\multicolumn{6}{|c|}{\textbf{Subject 64} (Record 12, Measured channels: C3, Cz and C4)}\\
\hline
%& \multicolumn{2}{|c|}{Frequencies in Hz}\\
Chnl & Delta & Theta & Alpha & Beta & Gamma\\

      &    (O) \hspace{3 mm}   (R)  &    (O) \hspace{3 mm}    (R)  &    (O)\hspace{3 mm}     (R)  &   (O)\hspace{3 mm}      (R)  &   (O)\hspace{2 mm}     (R)  \\  
\hline
  C5  &    0.441 0.444 &  0.137 0.131 &  0.080 0.070 &  0.194 0.200 &  0.061 0.060\\
  C1  &    0.431 0.441 &  0.135 0.136 &  0.079 0.070 &  0.208 0.209 &  0.055 0.066\\
  C2  &    0.435 0.437 &  0.139 0.136 &  0.076 0.072 &  0.208 0.217 &  0.057 0.066\\
  C6  &    0.452 0.421 &  0.110 0.116 &  0.058 0.062 &  0.162 0.176 &  0.052 0.053\\
\hline
Avg error & 2.6\%    &  3.2\%      &    8.5\%     &     4.0\%    &    10.5\%\\
%\multicolumn{6}{|c|}{Average percentage error in delta band: 0.837}\\
\hline
\end{tabular}
\end{table*}

\captionsetup{width=.8\textwidth}
%\end{landscape}

%\newpage
%\begin{figure*}[h!]
%	\centering
%	\includegraphics[width=0.5\textwidth]{electrodes.jpg}
%	\caption{Reconstructed signal and pseudospectrum}
%	\label{fig:plots1}	
%\end{figure*}

\end{document}